\documentclass[%reprint,
superscriptaddress,
nofootinbib,
 amsmath,amssymb,
]{revtex4-1}
\usepackage{amsmath}
\usepackage{amssymb}
\usepackage{bm}
\usepackage{color,graphicx}
\usepackage{slashed}
\usepackage{comment}
\newcounter{comment}

\begin{document}

\title{Transverse Orbital Angular Momentum in the Proton}

\author{Osamah Alkasassbeh}
\email{osaphy@mutah.edu.jo}
\affiliation{Mutah University - Department of Physics, Mutah, Alkarak - Jordan}
%\affiliation{New Mexico State University - Department of Physics, Box 30001 MSC 3D, Las Cruces NM, 88003 - USA}

\author{Abha Rajan}
\altaffiliation{Present address: NBC Framework, H.No-555 Ambala Rd., Sector 20, Huda, Kaithal Haryana, India -136027}
\email{ar5xc@virginia.edu}
\affiliation{Old Dominion University, 2100D Physical Sciences Bldg., Norfolk, VA 23529 - USA}

\author{Michael Engelhardt}
\email{engel@nmsu.edu}
\affiliation{New Mexico State University - Department of Physics, Box 30001 MSC 3D, Las Cruces, NM 88003 - USA}

\author{Simonetta Liuti}
\email{sl4y@virginia.edu}
\affiliation{University of Virginia - Physics Department,
382 McCormick Rd., Charlottesville, VA 22904 - USA} 

\vspace{1.5cm}

%\affiliation{$^1${Mutah University - Department of Physics, Mutah, Alkarak - Jordan}
%}

%\affiliation{$^2$ {Old Dominion University, 2100D Physical Sciences Bldg., Norfolk, VA 23529 - USA}
%}

%\affiliation{$^3${New Mexico State University - Department of Physics, Box 30001 MSC 3D, Las Cruces, NM 88003 - USA}
%}

%\affiliation{$^4$ {University of Virginia - Physics Department, 382 McCormick Rd., Charlottesville, VA 22904 - USA} 
%}

% \pacs{13.60.Hb, 13.40.Gp, 24.85.+p}

\begin{abstract}
Using the equations of motion of QCD and Lorentz invariance relations, we show a new, more direct way of obtaining the decomposition of the proton's quark transverse angular momentum into its spin and orbital components. The new decomposition can be understood as a twist-three relation, with the quark transverse spin described by the parton distribution $g_T$. We find that the orbital component can be expressed in terms of a moment in the quark transverse momentum, $k_T$, of the generalized transverse momentum-dependent distribution $F_{12}$, or alternatively in terms of the twist-three generalized parton distributions $H_{2T}$ and $\widetilde{E}_{2T}$.
\end{abstract}

\maketitle
%%%%%%%%%%%%%%%%%%%%%%
\noindent{\bf 1. Introduction} 
\vspace{0.2cm}

%%%%%%%%%%%%%%%%%%%%%%
\noindent Understanding the spin structure of the proton
as it emerges from the combined spin and orbital
angular momentum (OAM) of its quark and gluon constituents has been a central question in QCD, motivating current experimental and theoretical programs worldwide, including the future Electron Ion Collider (EIC). An extensive amount of literature has been dedicated to addressing the orbital component, as discussed, {\it e.g.}, in
\cite{Harindranath:1998ve,Burkardt:2008ua,Wakamatsu:2010cb,Hatta:2011ku,Ji:2020ena,Leader:2013jra,Ji:2012sj,Lorce:2011kd,Lorce:2011ni,Ji:2020hii,Bhattacharya:2022vvo}.

A QCD-based decomposition of the proton angular momentum into its quark and gluon spin and OAM components was first proposed by Jaffe and Manohar (JM) in  1990, \cite{Jaffe:1989jz}. 
In this early derivation, the authors could not, however, point at specific observables for the OAM components for the quark and gluon fields.
Subsequently, Ji suggested  
that the total angular momentum defined through the off-forward proton matrix elements of the QCD angular momentum tensor could be connected to the second Mellin moments of the Generalized Parton Distributions (GPDs) $H$ and $E$ \cite{Ji:1996ek}. 
GPDs enter the observables for deeply virtual exclusive scattering processes \cite{Ji:1996nm} and can be, in principle, extracted from experiment. 
Electron scattering provides, in fact, an accessible probe of the QCD angular momentum tensor's matrix elements rather than resorting directly to the gravitational interaction. 
Within Ji's framework the quark OAM part, $L$, is not obtained directly, but inferred by subtraction of the quark spin, $S$, from the total quark angular momentum, $J$. 
It was only after the work in \cite{Lorce:2011kd} that the quark OAM distribution in $x$ could be identified with a $k_T$ moment of a generalized transverse momentum distribution (GTMD), labeled $F_{14}$.\footnote{We adopt the naming conventions for GPDs and GTMDs given in Ref.~\cite{Meissner:2009ww} throughout this paper.}  
%%
%%%
A previous derivation using the Operator Product Expansion (OPE), in \cite{Polyakov:2002wz,Kiptily:2002nx}, showed that the difference, $J-S$, coincided with the second Mellin moment of a twist-3 GPD.
The two representations, through the GTMD $F_{14}$ and the twist-three GPD, were shown to be connected to one another as a consequence of the QCD equations of motion (EoM) and Lorentz invariance in Refs.~\cite{Rajan:2016tlg,Raja:2017xlo} (see also Ref.~\cite{Ji:2012ba}).
By identifying the specific parton distributions corresponding to the quark $J$, $L$ and $S$, one can then validate whether the relation
$J = L + S$ holds experimentally
using three separate measurements. Similar progress in the gluon sector, including non-trivial aspects of the gauge invariance of the theory, is presented in detail in Ref.~\cite{Bhattacharya:2022vvo}.

All of the progress described above refers to the partons' helicity, or the projection of angular momentum along the direction of motion.  In the case of transverse angular momentum, discussed, e.g., in \cite{Bakker:2004ib,Leader:2011cr,Ji:2012vj,Hatta:2012jm,Leader:2012ar,Lorce:2018zpf,Ji:2020hii,Guo:2021aik,Lorce:2021gxs},
the quark transverse spin vector, $S_{trans} $, does not conform to a simple parton model-based picture that implies a point-like photon-quark interaction. In order to describe the quark transverse spin, one needs to introduce 
the twist-three function $g_T(x)$, thus directly involving QCD interactions \cite{Jaffe:1996zw}. 
The Lorentz transformation properties of transverse angular momentum are also more subtle:
in this case, either the longitudinal parton position or momentum coordinates are involved, and therefore axial symmetry is broken; the generators of boosts and the rotations around the transverse axes do not commute, at variance with the longitudinal OAM case, where the generators for rotations around the longitudinal axis ({\it i.e.} in the transverse plane) commute with boosts \cite{Soper:1976jc}.  

In this paper we extend the method used to evaluate the decomposition of quark longitudinal angular momentum into its orbital and spin components in Refs.~\cite{Rajan:2016tlg,Raja:2017xlo} to the transverse component of the Pauli-Lubanski vector $W_{trans} $, namely, we derive the relation $W_{J,trans} = W_{L,trans} +W_{S,trans}$, 
\begin{eqnarray}
\label{eq:transverse_spin_SR}
\frac{1}{2} \int_0^1 dx\, x (H+E) & = &    \int_0^1 dx x \left( 
H + E+ \widetilde{E}_{2T}+ \lim_{\xi \rightarrow 0} \frac{H_{2T}}{\xi} \right) +  \frac{1}{2} \int_0^1 dx \, g_T \, \nonumber \\
W_{J,trans} \quad && \quad\quad\quad\quad\quad\quad\quad W_{L,trans} \quad\quad\quad \quad\quad\quad\quad\quad W_{S,trans}
\end{eqnarray}
applicable to the Ji decomposition of angular momentum; we also briefly address the extension to the JM decomposition. All GPDs in Eq.~(\ref{eq:transverse_spin_SR}) are taken in the forward limit.
We find that $W_{J,trans}$ is described by the same GPD combination, $H+E$, as the angular momentum $J_{long} $ in the longitudinal case, thus confirming the result of Ref.~\cite{Ji:2020hii}. On the other hand, the distribution $g_T=g_1+g_2$ characterizes transverse spin \cite{Jaffe:1990qh}: it is obtained by constructing the matrix element between proton states of the transverse component of the spin operator. $W_{L,trans}$ involves the twist-three GPDs $H_{2T}$ and $\widetilde{E}_{2T}$. It differs from longitudinal OAM \cite{Rajan:2016tlg,Raja:2017xlo} by a contribution $\lim_{\xi \rightarrow 0} (H_{2T}/\xi)$.\footnote{$H_{2T} $ is $\xi$-odd; it appears in the QCD correlation function/observables divided by $\xi$, therefore contributing in the $\xi \rightarrow 0$ limit defining the angular momentum sum rule.}
Our approach \cite{Rajan:2016tlg,Raja:2017xlo} uses the QCD EoM and Lorentz invariance relations (LIRs) to connect twist-three GPDs with the $k_T^2$-moments of GTMDs in the specific polarization configuration that defines transverse spin, thus involving a helicity flip in the correlation function: it is the proton helicity flip that generates the extra contribution from $H_{2T}/\xi$.

In what follows we expound the physics behind the new relation found in Eq.~(\ref{eq:transverse_spin_SR}),  describing the quark contributions to transverse proton spin.

\newpage
\noindent
{\bf 2.} {\bf Lorentz transformation properties of quark angular momentum}

\vspace{0.2cm}
\noindent 
The total quark angular momentum in the Ji decomposition of
the proton spin can be characterized through two mutually related quantities, both derived from
the quark angular momentum tensor
\begin{equation}
M_{\lambda \rho } = \int d V^{\alpha }
(r_{\lambda } T_{\alpha \rho }^{q} - r_{\rho } T_{\alpha \lambda }^{q} ) \ ,
\label{amtensor}
\end{equation}
constructed using the quark part of the Belinfante energy-momentum tensor $T_{\mu \nu }^{q} $.
Here, the 3-volume element normal vectors $dV^{\alpha } $ are taken to point in
the time direction in the rest frame of the proton, i.e., $dV^{\alpha } = \delta^{\alpha }_{0} d^3 r$ in that case; the integration
volume is subject to the relativity of simultaneity under Lorentz transformations. 
An important specification that must be emphasized is the reference point with respect to which the coordinates $r_{\mu } $ are defined. Here, we choose as the reference point the rest frame center of mass of the proton.
This guarantees that angular momentum transforms as a Lorentz tensor, since 
it is defined with respect to a reference point that itself transforms as a position vector. 
Other choices, and their behavior under
Lorentz boosts, were discussed in Refs.~\cite{Burkardt:2005hp,Lorce:2018zpf,Lorce:2021gxs}.

Based on Eq.(\ref{amtensor}) one can, on the one hand, consider the components of quark angular momentum $J$ itself,
\begin{equation}
J^i = \frac{1}{2} \, \epsilon^{0 i \lambda \rho } M_{\lambda \rho } ,
\label{jfromemt}
\end{equation}
where $i$ denotes a spatial component, and on the other hand, the components of the associated
Pauli-Lubanski vector $W_{J} $,
\begin{equation}
W_J^{i} = \frac{1}{2M} \, P_{\mu } \epsilon^{\mu i \lambda \rho } M_{\lambda \rho }  .
\label{wfromemt}
\end{equation}
where $M$ denotes the proton mass.
Therefore, the components of angular momentum $J^i $ transform as the (0 $i$)
components of a second rank tensor, Eq.~\eqref{jfromemt}, whereas the components of
the Pauli-Lubanski vector transform as the components of a vector, Eq.~\eqref{wfromemt}.
It should be emphasized that Eq.~(\ref{wfromemt}) represents the quark contribution
to the Pauli-Lubanski vector of the proton, as opposed to the
Pauli-Lubanski vector of the quark subsystem (which would contain
a factor $P^q $, the quark momentum operator, instead of $P$, the proton momentum), cf.~also the discussion in Ref.~\cite{Leader:2012ar}.

Discussions of longitudinal angular momentum usually focus on the angular momentum operator, $J$:
in addition to its prominent, direct significance as the generator of spatial rotations, its longitudinal component, $J_{long} $, is invariant under boosts. On the other hand, the longitudinal component of the Pauli-Lubanski vector, $W_{J,long}$, acquires a Lorentz $\gamma $-factor. 
In the case of transverse angular momentum,  
it is, instead, $W_{J,trans}$ that is invariant under boosts, whereas the transverse component of angular momentum, $J_{trans} $, acquires a Lorentz $\gamma$-factor. In view of this, we opt to focus the following discussion on the description in terms of the transverse Pauli-Lubanski vector; the alternative decomposition in terms of the angular momentum vector is given in the appendix.

In the proton rest frame, $J^i$ and $W_J^{i}$, $(i=1,2,3)$, coincide, so that one has, e.g., in the $3$-direction,
\begin{equation}
W_{J,0}^{3} = J_{0}^{3}
\label{eq:JW_rest}
\end{equation}
where the subscript $0$ labels these as the rest frame quantities.

Consider now a proton polarized in the $3$-direction, and observing it either in a frame boosted in the $3$-direction (corresponding to longitudinal polarization) or a direction perpendicular to the $3$-direction (corresponding to transverse polarization).
The Lorentz tensor structures imply that, in the former case, the $3$-component of angular momentum behaves as
\begin{eqnarray}
J^3_{long} &=& J^{3}_{0} 
\label{eq:J0J3}
\end{eqnarray}
whereas, in the latter case, the $3$-component of the Pauli-Lubanski vector behaves as
\begin{eqnarray}
W_{J, trans}^3 &=& W_{J,0}^{3}
\label{eq:WT}
\end{eqnarray}
We reiterate that Eq.~\eqref{eq:J0J3} describes longitudinal angular momentum and Eq.~\eqref{eq:WT} describes transverse angular momentum, even if in both cases the proton spin is pointing in the $3$-direction.

Combining (\ref{eq:JW_rest}), (\ref{eq:J0J3}) and (\ref{eq:WT}), one arrives at
\begin{equation}
W_{J,trans}^3 = J^3_0 = J^3_{long} = \frac{1}{2} \int_{0}^{1} dx\, x (H+E)
\label{wjt}
\end{equation}
where the right-most identity is the standard Ji relation for longitudinal angular momentum; the GPD combination $(H+E)$ is taken in the forward limit.

On the other hand, the Pauli-Lubanski vector associated with quark transverse spin can be identified with $g_T$, the transverse spin distribution function \cite{Jaffe:1990qh}
(cf.~also Ref.~\cite{Hughes:1983kf} for an early review), as, cf.~Ref.~\cite{Hatta:2012jm},
\begin{equation}
W^{3}_{S,trans} = \frac{1}{2} \int_{0}^{1} dx\, g_T =
\left. \frac{1}{2} \int_{0}^{1} dx\, H^{\prime }_{2T}
\right|_{\Delta =0}
\label{wtminus1}
\end{equation}
where in the second identity, the identification in terms of the forward limit of a GPD is given, $g_T(x) \equiv H^{\prime }_{2T}(x,0,0)$, taken at vanishing skewness, $\xi=0$, and $t\equiv \Delta^{2} =0$.

In what follows, we adapt the technique developed in Refs.~\cite{Rajan:2016tlg,Raja:2017xlo} for the longitudinal angular momentum case, using the QCD EoM and LIR to derive the GPDs and GTMDs for the transverse orbital angular momentum component described through the Pauli-Lubanski vector, $W^3_{L,trans}$. As already noted above, the following treatment will focus primarily on the Ji decomposition of proton spin; the extension to the JM counterpart will be addressed towards the end of the treatment.

\vspace{0.7cm}
\noindent{\bf 3. Lorentz Invariance Relations (LIR) and Equations of Motion (EoM) relations}
%%%%

\vspace{0.2cm}
\noindent LIR and EoM relations are obtained starting from the completely unintegrated correlation function \cite{Meissner:2009ww},
\begin{widetext}
\begin{equation}
 W^{\Gamma }_{\Lambda^{\prime } \Lambda }(P,k,\Delta) =
\frac{1}{2} \int \frac{d^4 z}{(2\pi )^4 } \, \, e^{ik\cdot z} 
\left\langle p', \Lambda^{\prime } \Big|
\bar{\psi } \left( -\frac{z}{2} \right) \Gamma \, {\cal U}\left( -\frac{z}{2} , \frac{z}{2} \right)
\psi \left( \frac{z}{2} \right) \Big|
p, \Lambda \right\rangle
\label{fund_corr}
\end{equation}
%%%

\vspace{0.2cm} 
\noindent where $p= P - \Delta/2 $, $p'= P + \Delta/2$,  $k=(k_{in}+k_{out})/2$ is the average quark four-momentum, and $\Delta = p' -p$. $\Lambda (\Lambda')$ is the helicity of the initial (final) proton.
$\Gamma $ stands for an arbitrary Dirac structure, namely, $\gamma^+, \gamma^+ \gamma_5, \sigma^{i+} \gamma_5$ for twist-two quantities, and $\gamma^i,\gamma^i\gamma_5, \sigma^{ij}\gamma_5$ (with $i,j$ transverse spatial indices) for twist-three quantities \cite{Mulders:1995dh,Tangerman:1994bb,Bacchetta:2006tn}.
\end{widetext}
One can consider two choices for the gauge link, ${\cal U}$, corresponding to the two formulations of the decomposition of angular momentum into its orbital and spin, quark and gluon components, derived by  Ji \cite{Ji:1996ek}, and by Jaffe and Manohar (JM) \cite{Jaffe:1989jz}, respectively. In the first case,  ${\cal U} $ runs along a straight path \cite{Ji:2012sj} between the
quark operators, while for JM a ``staple'' gauge link \cite{Hatta:2011ku} connects the quark fields to infinity and back \cite{Burkardt:2012sd}. 
Both correlators can be parametrized in terms of Lorentz-invariant amplitudes $A_i$, which, in the Ji case, are functions of the Lorentz invariants $k^2 , (k\cdot P), (k\cdot \Delta) , \Delta^{2} $.
In particular, an $A_i$-parametrization for the straight link case was constructed in \cite{Raja:2017xlo}, while the staple link case was given in \cite{Meissner:2009ww}.
On the other hand, the integral over $k^-$,
\begin{equation}
W^{\Gamma }_{\Lambda^{\prime } \Lambda } (P,x,k_T ,\xi , \Delta_{T} ) =
\int dk^{-} \ W^{\Gamma }_{\Lambda^{\prime } \Lambda } (P,k,\Delta)
\label{unint2gtmd}
\end{equation}
i.e., the GTMD correlation function, defines GTMDs via the parametrizations given in \cite{Meissner:2009ww}; these
are equally applicable in the straight and staple link cases, with the gauge link dependence contained in the GTMDs.\footnote{Note that GTMDs in general, in the staple-link case, contain a T-even (real) part and a T-odd (imaginary) part. Throughout the present treatment, all mentions of staple-link GTMDs refer exclusively to their T-even parts; the T-odd parts will play no role. Likewise, for the quark-gluon-quark terms introduced starting with Eq.~(\ref{eq:eqofmotion}), exclusively their T-even parts are referenced throughout in the staple link case. Of course, in the straight-link case, the distinction doesn't arise.}
$\Delta_{T} $ denotes the transverse momentum transfer. GPDs are obtained by further integrating (\ref{unint2gtmd}) over the transverse momentum $k_T $ \cite{Meissner:2009ww}.
    
As shown in Refs.~\cite{Raja:2017xlo,Rajan:2016tlg}, applying the QCD Equations of Motion (EoM), $(i \slashed{D}- m)\psi=0$, to the correlation function (\ref{unint2gtmd})
yields the following relation among the GTMD correlators,
\begin{eqnarray}
-\frac{\Delta^+}{2} W^{\gamma^i\gamma^5}_{\Lambda' \Lambda } +
ik^+\epsilon^{ij}W^{\gamma^j}_{\Lambda' \Lambda } +\frac{\Delta^{i} }{2}
W^{\gamma^+\gamma^5}_{\Lambda' \Lambda } -i \epsilon^{ij}k^j
W^{\gamma^+}_{\Lambda' \Lambda } + \mathcal{M}^{i, S}_{\Lambda' \Lambda } = 0
\label{eq:eqofmotion}
\end{eqnarray}
Here, $\mathcal{M}^{i, S}_{\Lambda' \Lambda } $  is the quark-gluon-quark ({\em qgq}) interaction term
resulting from the
derivative acting on the gauge link ${\cal U} $ upon integration of the equation of motion
by parts \cite{Raja:2017xlo}.\footnote{Here and in the following, the nomenclature of the quark-gluon-quark terms $\mathcal{M}^{i, S}_{\Lambda' \Lambda } $, $\mathcal{M}^{i, A}_{\Lambda' \Lambda } $, ${\cal M}_{F_{12}} $, ${\cal M}_{F_{14}} $, ${\cal M}_{G_{12}} $, etc. follows the definitions given in \cite{Raja:2017xlo}.} These off-forward relations, derived following the procedure first devised for transverse momentum distributions (TMDs) in Refs.~\cite{Mulders:1995dh,Tangerman:1994bb,Bacchetta:2003vn},
generalize the TMD relations given there. The detailed form of $\mathcal{M}^{i, S}_{\Lambda' \Lambda } $ is presented in \cite{Raja:2017xlo}; for the present discussion, note merely that in the straight-link
case, $\mathcal{M}^{i, S}_{\Lambda' \Lambda } $ vanishes identically when integrated
over all momentum components, $k_T$ as well as $x$. 
The relations in Eq.~\eqref{eq:eqofmotion} can be applied to either longitudinal, $\Lambda=\Lambda'$, or transverse, $\Lambda= -\Lambda'$, target polarization configurations. 
Using Eq.~(\ref{eq:eqofmotion}) for the transverse proton polarization configurations multiplied by $(\Delta_1 \pm i\Delta_2)$,  
\begin{equation}
(\Delta_1+i\Delta_2)W^\Gamma_{+-}+(\Delta_1-i\Delta_2)W^\Gamma_{-+}    ,
\label{eq:eom_phase}
\end{equation}
and inserting the parametrizations in terms of GTMDs/GPDs \cite{Meissner:2009ww} yields the EoM relation
\begin{eqnarray}
&& \!\! \left[H_{2T}' - \frac{\xi}{1-\xi^2} \left( \xi E_{2T}' - \widetilde{E}_{2T}' \right) \right]
\!+ \! x \left[\frac{1}{\xi} H_{2T} -\frac{1}{1-\xi^2} \left( \xi E_{2T} -
 \widetilde{E}_{2T} \right) \right] 
\!+ \frac{\Delta_T^2}{4M^2(1-\xi^2)} \widetilde{E}  \nonumber \\
&+& \frac{1}{2}\Big[ \frac{1}{\xi} F_{12}^{(1)} + \frac{\Delta_T^2}{2M^2( 1-\xi^2)} F_{14}^{(1)} \Big] 
+  \, i \frac{1}{4\xi \sqrt{1-\xi^{2}}} \, {\cal M}_{F_{12} } = 0  .
\label{eq:eom_transverse_long}
\end{eqnarray}
where the GTMD moment $F_{12}^{(1)} $ is defined as
\begin{equation}
F_{12}^{(1)} (x,\xi,\Delta_{T} ) = 2 \int d^2 k_T \, \frac{k_T^2}{M^2} \, \left[ \frac{k_T^2 \Delta_{T}^{2} - (k_T \cdot \Delta_{T} )^2 }{k_T^2 \Delta_{T}^{2} }
\right] F_{12} (x, k_T, \xi,\Delta_{T} )
\stackrel{\Delta_{T} \rightarrow 0}{\longrightarrow }
\int d^2 k_T \, \frac{k_T^2}{M^2} \, F_{12} (x, k_T, \xi,0)
\end{equation}
and analogously for $F_{14}^{(1)} $.
To obtain Eq.~\eqref{eq:eom_transverse_long}, one contracts Eq.~\eqref{eq:eom_phase} with
$\Delta^{i} /(4M\xi \Delta_{T}^{2} \sqrt{1-\xi^{2} })$, and finally integrates
over $k_T $. As mentioned above, the quark-gluon-quark term ${\cal M}_{F_{12} } $ is defined in Ref.~\cite{Raja:2017xlo}.

In Eq.~(\ref{eq:eom_transverse_long}), we have identified $k_T $-integrals of GTMDs  with GPDs, as given in Ref.~\cite{Meissner:2009ww}. It should be noted that, depending on the renormalization scheme, this identification of $k_T $-integrals of GTMDs with GPDs receives corrections analogous to the ones incurred when connecting TMDs to PDFs \cite{Boussarie:2023izj}. These renormalization scheme-dependent issues will not be pursued further here. Alternatively, the question of regulating the EoM relation can be addressed fully at the GTMD level 
in a chosen scheme, if one considers the GTMD
relation obtained before identifying $k_T $-integrals of GTMDs with GPDs to arrive at (\ref{eq:eom_transverse_long}).

LIRs are based on the covariant decomposition of the 
completely unintegrated quark-quark correlation function
(\ref{fund_corr}) appearing on the {\it rhs} of Eq.~(\ref{unint2gtmd}): 
the number of independent functions $A_i$ that parametrize this correlator in the straight-link case is less than the total number of
GTMDs parametrizing the {\it lhs} of Eq.~(\ref{unint2gtmd}); therefore relations among GTMDs arise. 
For the case presented here, the straight-link correlation function on the {\it rhs} of Eq.~(\ref{unint2gtmd}) for $\Gamma = \gamma^{\mu } $ is parametrized by eight Lorentz-invariant amplitudes $A_i$, as given in \cite{Raja:2017xlo}, whereas the correlation function on the {\it lhs} of Eq.~(\ref{unint2gtmd}) is parametrized  by four twist-two GTMDs (for $\Gamma = \gamma^{+} $) and eight twist-three GTMDs (for $\Gamma =\gamma^{i} $, with $i$ a transverse spatial index) \cite{Meissner:2009ww}.\footnote{There are another 4 twist-4 GTMDs for $\Gamma =\gamma^{-} $, which however do not play a role in this context.} 
In particular, inserting the respective parametrizations on either side of Eq.~(\ref{unint2gtmd}), one obtains the following LIR for the straight-link case,
\begin{equation}
\frac{\xi (1-\xi^{2}
 )}{2} \frac{d}{dx} \left(  \frac{1}{\xi} F_{12}^{(1)} +
\frac{\Delta_{T}^{2} }{2M^2 (1-\xi^{2} )} F_{14}^{(1)} \right) =
-(1-\xi^{2} )H_{2T} - \xi (H+E+\widetilde{E}_{2T} -\xi E_{2T} ) \ .
\label{eq:lir_transverse_long}
\end{equation}
It should again be noted that $k_T $-integrals of GTMDs have been identified with GPDs in arriving at the final
form of the LIR (\ref{eq:lir_transverse_long}), and therefore analogous caveats apply as in the case of the EoM. In addition, the LIR hinges on Lorentz invariance, and the adoption of a renormalization scheme that breaks Lorentz invariance will likewise engender corrections.

Eq.~\eqref{eq:lir_transverse_long}, in the forward limit and integrated over $x$, yields,
\begin{eqnarray}
 \frac{1}{2}
\int_0^1 dx \ \lim_{\xi \rightarrow 0} \frac{F_{12}^{(1)} }{\xi } = \int_0^1 dx\, x \left( \widetilde{E}_{2T}
+H+E+\lim_{\xi \rightarrow 0} \frac{H_{2T}}{\xi} \right)
\label{eq:f12rule}
\end{eqnarray}
Furthermore, inserting \eqref{eq:lir_transverse_long} into Eq.~\eqref{eq:eom_transverse_long}, integrating over $x$ (in which case the quark-gluon-quark term vanishes), and taking the forward limit, $\xi \rightarrow 0$, $\Delta_T \rightarrow 0$, we obtain our main result, Eq.~(\ref{eq:transverse_spin_SR}),
\begin{eqnarray}
\frac{1}{2}\int_0^1 dx \, x(H+E) &=& \int_0^1 dx\, x \left( \widetilde{E}_{2T}
+H+E+\lim_{\xi \rightarrow 0} \frac{H_{2T}}{\xi} \right)
+\frac{1}{2} \int_0^1 dx \, H^{\prime }_{2T}
\label{eq:h2toverxirule} 
\end{eqnarray}
Comparing Eq.~(\ref{eq:h2toverxirule}) with the quark contribution to the transverse Pauli-Lubanski vector of the proton, we find a direct correspondence between the {\it lhs} and the $J$ component of the transverse Pauli-Lubanski vector, Eq.~(\ref{wjt}), as well as between the second term on the {\it rhs} and the spin $S$ component, Eq.~(\ref{wtminus1}). 
One therefore can read off the transverse orbital, $L$, part as,
\begin{subequations}
\label{eq:wtlf12}
\begin{eqnarray}
W_{L,trans}^3 &=& \int_0^1 dx \, x \left( \widetilde{E}_{2T}
+H+E+\lim_{\xi \rightarrow 0} \frac{H_{2T}}{\xi} \right) \\
%%%
&=& \frac{1}{2}
\int_0^1 dx \ \lim_{\xi \rightarrow 0} \frac{F_{12}^{(1)} }{\xi }
\label{eq:wtlf12b}
\end{eqnarray}
\end{subequations}
To understand the physical reach of this formulation we juxtapose it to our previous finding for longitudinal OAM \cite{Rajan:2016tlg,Raja:2017xlo}, in Ji's definition of these quantities \cite{Ji:1996ek} (see Ref.~\cite{Raja:2017xlo} for details on the connection to the Jaffe-Manohar decomposition),
\begin{subequations}
\label{eq:Longitudinal}
\begin{eqnarray}
\frac{1}{2} \int_0^1 dx\, x (H+E) &=& \int_0^1 dx\, x (\widetilde{E}_{2T} +H+E)
\ \ + \ \ \frac{1}{2} \int_0^1 dx\, \widetilde{H}
\label{eq:gpddecomp} \\
%%%
&=& \hspace{0.9cm} -\int_0^1 dx\, F_{14}^{(1)} \hspace{0.9cm} + \ \
\frac{1}{2} \int_0^1 dx\, \widetilde{H}   
\label{eq:gtmddecomp} \\
\nonumber \\
J^3_{long} \hspace{1cm} &&  \hspace{1.6cm} L^3_{long} \hspace{1.6cm}
 \hspace{0.7cm} S^3_{long} \nonumber
\end{eqnarray}
\end{subequations}
(where, again, all distribution functions are taken in the forward limit).
Contrasting the transverse Pauli-Lubanski vector sum rule (\ref{eq:h2toverxirule}) with the longitudinal angular momentum sum rule (\ref{eq:gpddecomp}), one can make the following observations:
%%%%
%%%%
\begin{itemize}
\item As is well-known, the integrated spin contribution is the same for the longitudinal and transverse cases, $S^3_{long} = W^3_{S,trans} $. The integral of the longitudinal quark spin distribution encoded in $\widetilde{H}(x,0,0) = g_1(x)$ is the same as the integral of the transverse quark spin distribution encoded in $H^{\prime }_{2T}(x,0,0) = g_T(x) = g_1(x) +g_2(x) $, since, by the Burkhardt-Cottingham (BC) sum rule, the integrated additional term vanishes, $\int_0^1 dx \, g_2(x) =0$.\footnote{The BC sum rule can also be found in our approach using LIR and EoM relations, as shown in Ref.\cite{Raja:2017xlo}.}
\item This pattern in the spin contribution is mirrored in the orbital contribution. Since both $S^3_{long} = W^3_{S,trans} $ as well as $J^3_{long} = W^3_{J,trans} $, Eq.~(\ref{wjt}), it follows that $L^3_{long} = W^3_{L,trans} $. Referring to Eqs.~(\ref{eq:gpddecomp}) and (\ref{eq:h2toverxirule}), one sees that, compared to the longitudinal case, the GPD combination encoding OAM $L^3_{long} $, i.e.,
$(\widetilde{E}_{2T}+H+E)$, is modified in the transverse quantity, $W^3_{L,trans} $, by the additional term $(\lim_{\xi \rightarrow 0} H_{2T} /\xi )$.
This contribution must therefore
vanish upon taking the relevant second Mellin moment,
\begin{equation}
\int_0^1 dx\, x \lim_{\xi \rightarrow 0} \frac{H_{2T} }{\xi } = 0 \ ,
\end{equation}
amounting to an orbital counterpart to the BC sum rule. 
\item Before integrating over $x$, the distributions $g_2 $ and $(x \lim_{\xi \rightarrow 0} H_{2T} /\xi )$ distort the transverse spin and orbital contributions, respectively, compared to their longitudinal counterparts. Both distributions, integrating to zero, must have a node in $x$. It should be noted, however, that the two distributions do not cancel each other identically in Eq.~(\ref{eq:h2toverxirule}) before integrating over $x$, since further terms that integrate to zero have ceased to appear in the integrated forms (\ref{eq:gpddecomp}) and (\ref{eq:h2toverxirule}), such as quark-gluon-quark terms. The relation between the distributions is, in the forward limit,
\begin{equation}
g_2 (x) + x \lim_{\xi \rightarrow 0} \frac{H_{2T} (x)}{\xi } = -\int_{x}^{1} dy \ \lim_{\xi \rightarrow 0} \frac{H_{2T} (y) }{\xi } - \lim_{\xi \rightarrow 0} \frac{i}{4\xi } {\cal M}_{F_{12} } (x) + {\cal M}_{F_{14} } (x) \ ,
\end{equation}
as can be found by combining the EoM and LIR relations found above with their longitudinal counterparts in Ref.~\cite{Raja:2017xlo}.
\item Comparing the GTMD moment forms for the transverse (\ref{eq:wtlf12b}), and longitudinal (\ref{eq:gtmddecomp}) orbital components of the sum rules, one can see that the helicity non-flip GTMD, $F_{14}$, in the longitudinal decomposition has been replaced by the helicity flip GTMD, $F_{12}$. Note that division by $\xi $ combined with the $\xi \rightarrow 0$ limit corresponds to taking the derivative with respect to $\xi $ at $\xi =0 $; heuristically, if one interprets $\xi $ as representing the Fourier conjugate variable to longitudinal position $r_L $, this is suggestive of a factor $r_L $ being extracted.
In addition, the moment with respect to transverse momentum $k_T $ is taken. This suggests an interpretation of this term in the transverse sum rule as a $\langle r_L \times k_T \rangle $ orbital contribution. Such a picture in terms of longitudinal position $r_L $, of course, must be approached with caution in view of the relativistic distortions in the longitudinal direction.
A $\langle r_T \times k_L \rangle $ counterpart, on the other hand, evidently is absent in the transverse Pauli-Lubanski vector for the choice of reference point adopted here. One thus obtains a markedly simpler form of the sum rule in terms of $W^3_{L,trans} $ than in terms of $L^3_{trans} $, discussed in the appendix.
\end{itemize}
%%%%

The derivation discussed so far was for the Ji-type  definition of parton angular momentum. The GTMD moment form
(\ref{eq:wtlf12b}) of the transverse orbital contribution also suggests the generalization of the orbital term $W^3_{L,trans} $ to a Jaffe-Manohar type definition, by replacing the straight gauge link ${\cal U} $ in $F_{12} $ by a staple-shaped gauge link, in analogy to the correspondence found in the longitudinal case with the GTMD $F_{14} $
\cite{Hatta:2011ku,Ji:2012sj,Burkardt:2012sd}.
Similar to the longitudinal case \cite{Raja:2017xlo}, we find,
\begin{equation}
W^1_{L,trans,JM} - W^1_{L,trans,Ji} = -gv^{-} \lim_{\xi \rightarrow 0} \frac{1}{2\xi } \ \frac{1}{2P^{+} } \int_{0}^{1} ds\, \mbox{Im}
\langle P, S_1 | \bar{\psi } (0) \gamma^{+} U(0,sv) F^{+2} (sv) U(sv,0) \psi (0) | P,S_1 \rangle
\end{equation}
where here, for the sake of comparison with Ref.~\cite{Raja:2017xlo}, the proton has been chosen to be polarized along the transverse $1$-axis, with $P$ along the $3$-axis, and where $v$ denotes the vector describing the staple leg in the staple-shaped gauge link, cf.~the detailed discussion in \cite{Raja:2017xlo}.

Beyond the interpretation of the Mellin moments of Eqs.~(\ref{eq:eom_transverse_long}) and (\ref{eq:lir_transverse_long}) in terms of the quark angular momentum sum rules (\ref{eq:f12rule}) and (\ref{eq:h2toverxirule}), it should be emphasized that Eqs.~(\ref{eq:eom_transverse_long}) and (\ref{eq:lir_transverse_long}) are valid generally as functions of momentum fraction $x$, as well as $\xi $ and $\Delta_{T} $. The GTMD representation of the orbital term $W^{3}_{L,trans} $, in view of its Wigner function interpretation (cf.~the discussion of the longitudinal case in \cite{Lorce:2011kd}), suggests the possibility of {\em defining}, more generally, a $x$-density $W^{3}_{L,trans} (x)$ by not integrating over $x$ in (\ref{eq:wtlf12b}), and analogously for $L^3_{long} (x)$ in (\ref{eq:gtmddecomp}), with the caveat that a partonic interpretation is not straightforward in view of the presence of the gauge link in the TMD operator. If one ventures to adopt such a definition, Eqs.~(\ref{eq:eom_transverse_long}) and (\ref{eq:lir_transverse_long}) can be viewed as statements about $W^{3}_{L,trans} (x)$ in the forward limit.

Finally, we derive $x$-dependent expressions for the twist-three GPDs describing transverse OAM through its Pauli-Lubanski identification Eq.~\eqref{eq:wtlf12}, in terms of twist-two GPDs. In the approximation where the $qgq$ term and the quark mass can both be disregarded, these amount to generalized Wandzura-Wilczek (WW) \cite{Wandzura:1977qf}
relations (see also the detailed discussion in \cite{Raja:2017xlo}). For the combination $\widetilde{E}_{2T} + \lim_{\xi \rightarrow 0} \frac{1}{\xi } H_{2T} $, we first obtain, in the forward limit,\footnote{To arrive at this result, the EoM and LIR (\ref{eq:eom_transverse_long}) and (\ref{eq:lir_transverse_long}) must be further supplemented by the EoM and LIR involving the GTMD moment $G_{12}^{(1)} $ from Ref.~\cite{Raja:2017xlo} (Eqs.~(119) and (132) therein), in order to eliminate $H_{2T}^{\prime } $.}
\begin{eqnarray}
\label{eq:WW_trans}
\widetilde{E}_{2T} + \lim_{\xi \rightarrow 0} \frac{1}{\xi } H_{2T} &=& -\int_{x}^{1} \frac{dy}{y} (H+E) - \int_{x}^{1} \frac{dy}{y^2 } \widetilde{H} 
+\frac{m_q}{M} \left[ -\frac{1}{x^2 } H_T + 2\int_{x}^{1} \frac{dy}{y^3 } H_T \right] + {\cal M}_{qgq} 
\end{eqnarray}
where $m_q$ is the quark mass, and $H_T$ is the transversity GPD with $H_T(x,0,0) = h_1(x)$ \cite{Diehl:2002he}. 
The (straight-link) $qgq$ term reads
\begin{eqnarray}
{\cal M}_{qgq} =   -\frac{1}{x^2 } {\cal M}_{G_{12}} + 2\int_{x}^{1} \frac{dy}{y^3 } {\cal M}_{G_{12}} -\frac{1}{x} \lim_{\xi \rightarrow 0} \frac{i}{4\xi } {\cal M}_{F_{12 } } + \int_{x}^{1} \frac{dy}{y^2 }
\lim_{\xi \rightarrow 0} \frac{i}{4\xi }
{\cal M}_{F_{12 } }
\end{eqnarray}
In the forward limit considered here, the quark-gluon-quark terms reduce to,
\begin{eqnarray}
{\cal M}_{G_{12}} &=& \int d^2 k_T \frac{1}{2M} \left( {\cal M}_{+-}^{1,A} + {\cal M}_{-+}^{1,A} \right) \ ,\\
{\cal M}_{F_{12} } &=& -i\ \int d^2 k_T \frac{1}{M} \left( {\cal M}_{+-}^{1,S} + {\cal M}_{-+}^{1,S} \right)
\end{eqnarray}
(equivalent forms using ${\cal M}^{2,\{A,S\} } $ can also be given).
The WW approximation is obtained dropping the quark mass and the $qgq$ terms,
\begin{equation}
\label{eq:WW_transverse}
\widetilde{E}_{2T} + \lim_{\xi \rightarrow 0} \frac{1}{\xi } H_{2T} = -\int_{x}^{1} \frac{dy}{y} (H+E) - \int_{x}^{1} \frac{dy}{y^2 } \widetilde{H} 
\end{equation}
Similarly, in Ref.~\cite{Raja:2017xlo} we found, in the longitudinal case, the following WW relation involving $\widetilde{E}_{2T}$,
\begin{eqnarray}
\widetilde{E}_{2T} =  -\int_{x}^{1} \frac{dy}{y} (H+E) - \frac{1}{x} \widetilde{H} + \int_{x}^{1} \frac{dy}{y^2 } \widetilde{H}
\label{eq:longWW}
\end{eqnarray}
Combining Eqs.~(\ref{eq:WW_transverse}) and (\ref{eq:longWW}) gives a WW-relation for $\lim_{\xi \rightarrow 0} \frac{1}{\xi } H_{2T}$ alone,
\begin{equation}
\lim_{\xi \rightarrow 0} \frac{1}{\xi } H_{2T} = \frac{1}{x} \widetilde{H} - 2 \int_{x}^{1} \frac{dy}{y^2 } \widetilde{H} 
\end{equation}
On the other hand, if one adopts the definitions for
the $x$-distributions characterizing longitudinal and transverse OAM suggested above, their WW-approximations are, respectively, given by,
\begin{subequations}
\begin{eqnarray}
L^3_{long} (x) &=& -F_{14}^{(1)}
= x \int_{x}^{1} \frac{dy}{y} (H+E) - x \int_{x}^{1} \frac{dy}{y^2 } \widetilde{H} \ , \\
W_{L,trans}^3(x) &=& \lim_{\xi \rightarrow 0} \frac{1}{2\xi } F_{12}^{(1)}
= x \int_{x}^{1} \frac{dy}{y} (H+E) - \int_{x}^{1} \frac{dy}{y} \widetilde{H} + x \int_{x}^{1} \frac{dy}{y^2 } \widetilde{H} \ .
\end{eqnarray}
\end{subequations}

\vspace{0.5cm}
\noindent {\bf 4. Summary and Outlook} 

\vspace{0.2cm}
\noindent 
We present a decomposition of the quark contribution $W^3_{J,trans} $ to the transverse Pauli-Lubanski vector of the proton into its orbital, $W^3_{L,trans} $, and spin $W^3_{S,trans} $ components, where, at variance
with previous derivations, all three quantities are fully specified in terms of parton distribution functions. The central relations (\ref{eq:h2toverxirule}),(\ref{eq:wtlf12})
bring transverse quark angular momentum within experimental grasp by allowing us to identify, for each component, the observables attainable in deeply virtual exclusive experiments.

An important element in the derivation was the choice of the rest frame center of mass as the reference point for the quark position vector $r$ entering the definition of the angular momentum tensor, guaranteeing simple Lorentz transformation properties for the transverse Pauli-Lubanski
vector, and allowing us to identify the GPD content of $W^3_{J,trans} $. Furthermore, the transverse spin component $W^3_{S,trans} $ is given by the twist-three function $g_T $. Complementing this with equation of motion (EoM) and Lorentz invariance (LIR) relations \cite{Rajan:2016tlg,Raja:2017xlo} allows us to
infer also the orbital component $W^3_{L,trans} $. It can be expressed in terms of the twist-three GPDs $\widetilde{E}_{2T} $ and $H_{2T} $, or,
alternatively, in terms of a $k_T $-moment of the GTMD $F_{12} $. We find that the GPD $H_{2T} $ satisfies an orbital counterpart of the Burkhardt-Cottingham sum rule.

Whereas the primary focus of the derivation lay on the Ji-type decomposition of the angular momentum into its orbital and spin components, the representation of the orbital part in terms of the GTMD $F_{12} $ suggests an extension of our results to the Jaffe-Manohar-type decomposition by utilizing the gauge link dependence of the GTMD; we exhibit the quark-gluon-quark term representing the difference between the two forms. The GTMD representation also suggests the definition of a density in momentum fraction $x$ of the orbital part, in view of its Wigner function interpretation. Finally, we
give Wandzura-Wilczek approximations to the various twist-three functions arising in our treatment.

The EoM/LIR method, originally developed in Refs.~\cite{Rajan:2016tlg,Raja:2017xlo} to derive
relations among measurable quark distributions describing longitudinal angular momentum, and extended here to the transverse angular momentum case, will, in further work, be more widely applied to the off-forward domain, including the chiral-odd sector.

\vspace{0.5cm}
\noindent {\bf Appendix} 

\vspace{0.2cm}
\noindent
This appendix exhibits the interpretation of the sum rules
(\ref{eq:h2toverxirule}),(\ref{eq:f12rule}) in terms of the quark angular momentum
$J$, complementing the discussion of the Pauli-Lubanski vector $W_J $ given above.
Starting from the rest frame angular momentum $J^3_0 $, cf.~Eq.~({\ref{wjt}), the Lorentz transformation properties of $J$ imply that, under a boost in a direction perpendicular to the 3-direction, $J^3 $ acquires a Lorentz $\gamma $-factor,
\begin{equation}
J^3_{trans} = \gamma J^3_0 = P^{+} \frac{1}{M \sqrt{2} } \frac{1}{2} \int_{0}^{1} dx\, x (H+E)
+\frac{1}{P^{+} } \frac{M}{2\sqrt{2} } \frac{1}{2} \int_{0}^{1} dx\, x (H+E)
\label{jtminus1}
\end{equation}
where the $\gamma $-factor $\gamma =P^0 /M$ has been organized in terms of powers of $P^{+} $.

On the other hand, one can express quark transverse spin through the transverse spin distribution function
$g_T $. The transverse spin degree of freedom vanishes in the limit $P^{+} \rightarrow \infty $, and accordingly, referring to the definition of $g_T $ \cite{Jaffe:1991kp}
\begin{equation}
\frac{1}{2} \int \frac{d\lambda }{2\pi } e^{i\lambda x}
\langle P S | \bar{\psi } (0) \gamma_{\mu } \gamma_{5} \psi (\lambda n) | P S \rangle
= (S\cdot n) p_\mu g_1(x) + (S_T)_\mu g_T(x) + M^2 (S \cdot n) n_\mu  g_3(x),
\label{eq:Jaffespin}
\end{equation}
the quark transverse spin only acquires a
contribution proportional to $1/P^{+} $,
\begin{equation}
S^3_{trans} = \frac{1}{P^{+} } \frac{M}{\sqrt{2} } \ \frac{1}{2} \int_{0}^{1} dx\, g_T = \frac{1}{P^{+} }
\left. \frac{M}{\sqrt{2} } \ \frac{1}{2} \int_{0}^{1} dx\, H^{\prime }_{2T}
\right|_{\Delta =0}
\label{stminus1}
\end{equation}
Combining this input with the sum rules (\ref{eq:h2toverxirule}) and (\ref{eq:f12rule}), one can now complete the decomposition of transverse quark angular momentum. The following allocation of contributions results in view of Eqs.~(\ref{jtminus1})
and (\ref{stminus1}): Since there is no quark transverse spin contribution at order $P^{+} $, the transverse angular momentum is purely orbital at that order; on the other hand, at order $1/P^+$, one then has to re-balance the orbital terms to arrive at
\begin{eqnarray}
L^3_{trans} &=& P^{+} \frac{1}{M \sqrt{2} } \frac{1}{2} \int_{0}^{1} dx\, x (H+E) + \frac{M}{\sqrt{2}P^+ } \left[ \int_{0}^{1} dx \, x \left( \widetilde{E}_{2T}
+H+E+\lim_{\xi \rightarrow 0} \frac{H_{2T}}{\xi} \right)
-\frac{1}{4}\int_{0}^{1} dx\, x(H+E) \right] \\
&=& P^{+} \frac{1}{M \sqrt{2} } \frac{1}{2} \int_{0}^{1} dx\, x (H+E) + \frac{M}{\sqrt{2}P^+ } \left[  \frac{1}{2} \lim_{\xi \rightarrow 0}
\int_{0}^{1} dx \frac{F_{12}^{(1)} }{\xi }
- \frac{1}{4}\int_{0}^{1} dx\, x(H+E) \right]
\end{eqnarray}
Note that, for finite $P^{+} $, the $\int_{0}^{1} dx\, x (H+E)$ contributions tend to
cancel in $L^3_{trans} $, whereas they add in $J^3_{trans} $.
The specific allocation of orbital terms arrived at here can be viewed as being related to the specific choice of reference point for angular momentum; changing the reference point would add equal contributions to $J^3_{trans} $ and $L^3_{trans} $ on either side of the sum rule $J^3_{trans} = L^3_{trans} + S^3_{trans} $.

\vspace{0.5cm}
\noindent {\bf Acknowledgments}

\vspace{0.2cm}
\noindent
This research is supported by DOE grants DE-SC0016286 (S.L.) and DE-FG02-96ER40965 (M.E. and O.A.).

\bibliography{OAM_bib}

\end{document}